\documentclass{emulateapj}
\usepackage{amsmath,amssymb}
\usepackage{graphicx,mathptm} 
\usepackage{times} 
\usepackage{natbib}
\def\be{\begin{equation}}
\def\ee{\end{equation}}

\def\bdm{\begin{displaymath}}
\def\edm{\end{displaymath}}
\def\kr{cosmic ray }
\def\krs{cosmic rays }

\def\fp{Fokker-Planck }

\def\ppa{p_{\parallel }}

\def\kpa{k_{\parallel }}
\def\kper{k_{\perp }}
\def\a{\alpha }

\def\be{\begin{equation}}
\def\ee{\end{equation}}
\def\bdm{\begin{displaymath}}
\def\edm{\end{displaymath}}

\def\kr {cosmic ray }
\def\be {\begin{equation}}
\def\ee {\end{equation}}

\def\krs {cosmic rays }

\def\c\kn{critical Klein-Nishina Lorentz factor }
\def\omr{\omega _R}
\def\XX{\vec{X}}
\def\pa{\partial }
\def\fo{Fokker-Planck }
\def\vt{\vert }

\def\mgf{\vec{B}}
\def\eef{\vec{E}}

\def\ebe{\end{displaymath}\begin{equation}}
\def\eba{\end{displaymath}\begin{displaymath}}

\begin{document}
\shorttitle{Cosmic ray acceleration at relativistic shock waves}
\shortauthors{R. Schlickeiser}
\title{Diffusive cosmic ray acceleration at relativistic shock waves with magnetostatic turbulence}
\author{R. Schlickeiser}
\affil{Institut f\"ur Theoretische Physik, Lehrstuhl IV:
Weltraum- und Astrophysik, Ruhr-Universit\"at Bochum
D-44780 Bochum, Germany}
\email{rsch@tp4.rub.de}
\begin{abstract}
The analytical theory of diffusive cosmic ray acceleration at parallel stationary shock waves with magnetostatic turbulence is generalized to arbitrary shock speeds $V_s=\beta _1c$, including in particular relativistic speeds. This is achieved by applying the diffusion approximation to the relevant Fokker-Planck particle transport equation formulated in the mixed comoving coordinate system. In this coordinate system the particle's momentum coordinates $p$ and $\mu =p_{\parallel }/p$ are taken in the rest frame of the streaming plasma, whereas the time and space coordinates are taken in the observer's system. For magnetostatic slab turbulence the diffusion-convection transport equation for the isotropic (in the rest frame of the streaming plasma) part of the particle's phase space density is derived. For a step-wise shock velocity profile the steady-state diffusion-convection transport equation is solved. For a symmetric pitch-angle scattering Fokker-Planck coefficient $D_{\mu \mu }(-\mu )=D_{\mu \mu }(\mu )$ 
 the steady-state solution is independent of the microphysical scattering details. For nonrelativistic mono-momentum particle injection at the shock the differential number density of accelerated particles is a Lorentzian-type distribution function which at large momenta approaches a power law distribution function $N(p\ge p_c)\propto p^{-\xi }$ with the spectral index $\xi (\beta _1) =1+[3/(\Gamma _1\sqrt{r^2-\beta _1^2}-1)(1+3\beta _1^2)]$. For nonrelativistic ($\beta _1\ll 1$) shock speeds this spectral index agrees with the known result $\xi (\beta _1\ll 1)\simeq (r+2)/(r-1)$, whereas for ultrarelativistic ($\Gamma _1\gg 1$) shock speeds the spectral index value is close to unity.
    \end{abstract}
\keywords{acceleration of particles -- cosmic rays -- relativity -- shock waves }
%
%
%
\section{Introduction}
One of the most important problems of modern astrophysics is to explain how cosmic ray particles
are accelerated to relativistic energies in powerful sources of nonthermal radiation. Diffusive first-order Fermi acceleration at nonrelativistic shock fronts has been regarded as a prime candidate for particle acceleration in astrophysics (for reviews see Drury 1983; Blandford and Eichler 1987). Modern TeV air-Cherenkov telescopes have indeed resolved the shock regions in supernova remnants and
identied the shocks as strong emission regions of TeV photons generated by the accelerated particles
(Hinton and Hofmann 2009). 

It is likely that diffusive shock acceleration also operates efficiently at magnetized shock waves with relativistic speeds. Such relativistic shocks form during the interaction of relativistic supersonic and super-Alfvenic outflows with the ambient ionized interstellar or intergalactic medium (Gerbig and Schlickeiser 2009) producing anisotropic counterstream plasma distribution functions due to shock-reflected charged particles in the upstream medium (Spitkovsky 2008, Sironi and Spitkovsky 2009). Relativistic outflows are a direct consequence of violent explosive events such as in gamma-ray burst sources (Piran 1999) both in the collapsar (Woosley 1993, Paczynski 1998) and supranova (Vietri and Stella 1998) models, but also occur as highly collimated pulsar winds and jets of active galactic nuclei with initial bulk Lorentz factors $\Gamma _0=(1-(V_0/c)^2)^{-1/2}\simeq 400$. 

The transport and acceleration of energetic particles in the partially turbulent cosmic magnetic fields associated with shocks is
described using the \fo  equation for the particle distribution function (for a recent derivation, see Schlickeiser 2011). The diffusion approximation for the particle density in the rest frame of the fluid is a well-known simplied form of the \fo equation, which results when turbulent pitch-angle scattering is strong enough to ensure that the scale of the particle density variation is signicantly greater than the
particle mean free path (Jokipii 1966, Hasselmann and Wibberenz 1968, Earl 1974). Numerical studies confirmed the accuracy of the diffusion approximation in a uniform mean guide magnetic field (Kota et al. 1982).

While for nonrelativistic shock waves the analytic theory of diffusive shock acceleration is well developed (Axford et al. 1977, Krymsky 1987, Blandford and Ostriker 1978, Bell 1978, Drury 1983), for relativistic shock speeds such an analytical theory does not exist sofar even for parallel shock waves, although the underlying \fo transport equation (see Eq. (\ref{a3}) below) for the particle dynamics has already been derived (Webb 1985; Kirk, Schneider and Schlickeiser 1988) some years ago. The existing literature concentrated on semi-numerical eigenfunction solutions of this Fokker-Planck transport equation pioneered by Kirk and Schneider (1987), relativistic Monte Carlo simulations (Ellison et al. 1990, Ostrowski 1991, Bednarz and Ostrowski 1998, Summerlin and Baring 2012), and relativistic particle in-cell simulations (Spitkovsky 2008, Sironi and Sptkovsky 2009). It is the purpose of this manuscript to develop an analytical study of \kr acceleration in parallel relativistic magnetized shock waves 
employing the diffusion approximation in the upstream and downstream regions of the shock wave. The development runs much in parallel with the existing work on nonrelativistic shocks. 
\section{Basic equations}
Magnetized space plasmas such as the interstellar medium harbour low-frequency linear ($\delta B\ll B_0$) transverse MHD waves (such as shear Alfven and magnetosonic plasma waves) with dispersion relations $\omr^2=V_A^2\kpa ^2$ and $\omr^2=V_A^2k^2$, respectively, in the rest frame of the moving plasma. Faraday$^{'}$s induction law then indicates for MHD waves that the strength of turbulent electric fields $\delta E=(V_A/c)\delta B \ll \delta B$ is much smaller than the strength of turbulent magnetic fields. The ordering $B_0\gg \delta B\gg \delta E$ corresponds to the derivation of \kr transport equations for $<f>(\vec{X}, p,\mu, \phi ,t)\to f_0(\vec{X}, p,\mu, t)\to F(\vec{X},p,t)$ from the collisionfree Boltzmann equation fir the full phase space distribution $<f>(\vec{X}, p,\mu, \phi ,t)$ to the \fo equation for its gyrotropic part $f_0(\vec{X}, p,\mu, t)$, and to the diffusion-convection transport equation for its isotropic part $F(\vec{X},p,t)$, respectively (for a recent review see  
Schlickeiser 2011). Accordingly, the \kr anisotropy, defined as the deviation 

\be
g(\XX,v,\mu ,t)=f_0(\vec{X}, p,\mu, t)-F(\vec{X},p,t),
\label{a1}
\ee
then is small ($\vt g \vt \ll F$) with respect to $F$. The diffusion approximation applied to the \fo transport equation for $f_0(\vec{X}, p,\mu, t)$ allows us to relate the \kr anisotropy $g$ to the solutions of the diffusion-convection tranport equation for $F$. 

Because of the gyrorotation of the \kr particles in the uniform magnetic field, one is not so much interested in their actual position as in the coordinates of the \kr guiding center 

\be
\vec{X}=(X,Y,Z)=\vec{x}+{c\over q_aB_0^2}\vec {p}\times \vec{B}_0=
\vec{x}+{c\over q_aB_0}\begin{pmatrix} p_y \\ -p_x \\ 0\end{pmatrix},
\label{a2}
\ee
where we orient the large-scale guide magnetic field, which is uniform on the scales of the \kr particles gyradii $R_L=v/\vert \Omega \vert $, $\vec{B}_0=B_0\vec{e}_z=(0,0,B_0)$ along the $z$-axis. $v$ and $\Omega _a=q_aB_0/\gamma m_ac$ denote the speed and the relativistic gyrofrequency of a \kr particle with mass $m$, charge $g_a$ and energy $\gamma m_ac^2$. 

The Larmor-phase averaged \fp transport equation in a medium, propagating with the stationary bulk speed $\vec{U}=U(z)\vec{e_z}$ 
with $\Gamma =[1-(U^2/c^2)]^{-1/2}$ aligned along the magnetic field direction is given by (Webb 1985; Kirk, Schneider and Schlickeiser 1988)

\bdm
\Gamma \left[1+{Uv\mu \over c^2}\right]{\pa f_0\over \pa t}+\Gamma \left[U+v\mu \right]{\pa f_0\over \pa z}+{v(1-\mu ^2)\over 2L}{\partial f_0\over \partial \mu }
\eba
-\a (z)\left(\mu +{U\over v}\right)\left[\mu p{\pa f_0\over \pa p}+(1-\mu ^2){\pa f_0\over \pa \mu }\right]
+{\cal R}f_0-S(\XX ,p,t)
\eba
={\pa \over \pa \mu }\left[D_{\mu \mu } {\pa f_0\over \pa \mu }+D_{\mu \sigma } {\pa f_0\over \pa y_{\sigma }}\right]
\ebe
+p^{-2}{\pa \over \pa y_{\omega }}p^2\left[D_{\omega \mu } {\pa f_0\over \pa \mu }+D_{\omega \sigma }{\pa f_0\over \pa y_{\sigma }}\right],
\label{a3}
\ee
irrespective of how the \fo coefficients are calculated, either by quasilinear (Schlickeiser 2002) or nonlinear (Shalchi 2009) 
\kr transport theories. $\a (z)$ denotes the rate of adiabatic deceleration/acceleration in relativistic flows 

\be
\a (z)={c^2\over U(z)}{d\Gamma (z)\over dz}={dU\over dz}\Gamma ^3={d(U\Gamma )\over dz}
\label{a4}
\ee
In the \fo Eq. (\ref{a3}) the phase space coordinates have to be taken in the mixed comoving coordinate system (time and space coordinates in the laboratory (=observer) system and particle's momentum coordinates $p$ and $\mu =\ppa /p$ in the rest frame of the streaming plasma). Moreover, in Eq. (\ref{a3}) we use the Einstein sum convention for indices, and $y_{\omega },y_{\sigma }\in [p,X,Y]$ represent the three phase space variables with non-vanishing stochastic fields $\delta \eef $ and $\delta \mgf $. Consequently, the term on the right-hand side generally represents 9 different \fp coefficients: but, depending on the turbulent fields considered, not all of them are non-zero and some are much larger than others. $S(\XX ,p,t)$ accounts for additional sources and sinks of particles.

The focusing length (Roelof 1969) 

\be
L=-{B_0^2\over \vec{B}_0\cdot \hbox{grad}\, (\vt \vec{B}_0\vt)}=-{B_0(z)\over (dB_0(z)/dz)}
\label{a5}
\ee
represents the spatial gradient of the guide magnetic field $B_0$. It is the only term, resulting from the mirror force in the large scale inhomogeneous guide magnetic field, that we consider (neglecting possible drift effects). $L>0$ for a diverging guide magnetic field; $L<0$ for a converging guide magnetic field.

\be
{\cal R}f=-p^{-2}{\pa \over \pa p}\left[p^2\dot{p}_{\rm loss}f\right]+{f\over T_c}
\label{a6}
\ee
represents continuous ($\dot{p}_{\rm loss}$) and catastrophic ($T_c$) momentum losses of \kr particles.

For spatially constant flows the rate of adiabatic deceleration/acceleration (\ref{a4}) vanishes,
 and the remaining flow velocity ($U$)  dependent terms in Eq. (\ref{a3}) simply result from the Lorentz transformation of special relativity of the laboratory-frame position-time coordinates  $(z,t)$ to the mixed-comoving-frame position-time coordinates $(z^{'},t^{'})$. However, for spatially varying flow speeds 
$U(z)$ special relativity no longer applies and has to be replaced by the transformation laws from general realtivity. As noted by 
Riffert (1986) as well as Kirk, Schneider and Schlickeiser (1988) these introduce connection coefficients or Christoffel symbols of the first kind. In a flat Euclidean space-time metric the terms proportional to $\a (z)$ in Eq. (\ref{a3}) are exactly these connection coefficients. 
\section{Magnetostatic slab turbulence}
As important special case we consider magnetostatic (vanishing turbulent electric fields), isospectral, slab ($\kper =0$) turbulence. Then the only nonvanishing \fo coefficient is $D_{\mu \mu }$. In this case the \fo transport equation (\ref{a3}) simplifies to 

\bdm
\Gamma \left[1+{Uv\mu \over c^2}\right]{\pa f_0\over \pa t}+\Gamma \left[U+v\mu \right]{\pa f_0\over \pa z}+{v(1-\mu ^2)\over 2L}{\partial f_0\over \partial \mu }
\eba
-\a (z)\left(\mu +{U\over v}\right)\left[\mu p{\pa f_0\over \pa p}+(1-\mu ^2){\pa f_0\over \pa \mu }\right]
+{\cal R}f_0
\ebe
=S(\XX ,p,t)+{\pa \over \pa \mu }\left[D_{\mu \mu } {\pa f_0\over \pa \mu } \right]
\label{b1}
\ee
Due to the rapid pitch angle scattering the gyrotropic particle distribution function $f_0(\vec{X},p,\mu ,t)$ adjusts very quickly to a distribution function which is close to the isotropic distribution in the rest frame of the moving background plasma. Defining the isotropic part of the phase space density $F(\vec{X},z,p,t)$ as the $\mu $-averaged 
phase space density 

\be
F(\vec{X},p,t)\equiv {1\over 2}\int_{-1}^1d\mu \; f_0(\vec{X},p,\mu ,t),
\label{b2}
\ee
we follow the analysis of Jokipii (1966) and Hasselmann and Wibberenz (1968) to split the total density $f_0$ into the isotropic part $F$ and an anisotropic part $g$,

\be
f_0(\XX ,p,\mu ,t)=F(\XX ,p,t)+\; g(\XX ,p,\mu ,t)
\label{b3}
\ee
where because of Eq. (\ref{b2}) 

\be
\int _{-1}^1d\mu \, g(\XX ,p,\mu ,t)=0
\label{b4}
\ee
Inserting the ansatz (\ref{b3}) yields for the \fo equation (\ref{b1}) 

\bdm
\Gamma \left[1+{Uv\mu \over c^2}\right]\left({\pa F\over \pa t}+{\pa g\over \pa t}\right)+\Gamma \left[U+v\mu \right]\left({\pa F\over \pa z}+{\pa g\over \pa z}\right)
\eba
-\a (z)\left(\mu +{U\over v}\right)\left[\mu p{\pa F\over \pa p}+\mu p{\pa g\over \pa p}+(1-\mu ^2){\pa g\over \pa \mu }\right]
\ebe
+{v(1-\mu ^2)\over 2L}{\partial g\over \partial \mu }+{\cal R}F+{\cal R}g-S(\XX ,p,t)
={\pa \over \pa \mu }\left[D_{\mu \mu } {\pa g\over \pa \mu }\right]
\label{b5}
\ee
Averaging this equation over $\mu $, using that $D_{\mu \mu }(\mu =\pm 1)=0$, leads to the diffusion-convection transport equation 

\bdm
\Gamma \left[{\pa F\over \pa t}+U{\pa F\over \pa z}\right]-{\a \over 3}p{\pa F\over \pa p}+{\cal R}F-S(\XX ,p,t)
\eba
+{v\over 2}\left[\Gamma \left({\pa \over \pa z}+{U\over c^2}{\pa \over \pa t}\right)+{1\over L}\right]
\int_{-1}^1d\mu \, \mu g
\ebe
-{\a U\over 2v}\left[p{\pa \over \pa p}+2\right]\int_{-1}^1d\mu \, \mu g-{\a \over 2}\left[p{\pa \over \pa p}+3\right]\int_{-1}^1d\mu \, \mu ^2g
=0,
\label{b6}
\ee
involving the first and second moment of the anisotropy. 

Subtracting Eq. (\ref{b6}) from the \fo equation (\ref{b5}) provides 

\bdm
\Gamma v\mu \left({\pa F\over \pa z}+{U\over c^2}{\pa F\over \pa t}\right)+\Gamma v\mu \left({\pa g\over \pa z}+{U\over c^2}{\pa g\over \pa t}\right)
+\Gamma \left({\pa g\over \pa t}+U{\pa g\over \pa z}\right)
\eba
+{\a \over 3}p{\pa F\over \pa p}+{\cal R}g-\a \left(\mu +{U\over v}\right)\left[\mu p{\pa F\over \pa p}+\mu p{\pa g\over \pa p}+(1-\mu ^2){\pa g\over \pa \mu }\right]
\eba
-\left[{v\Gamma \over 2}\left({\pa \over \pa z}+{U\over c^2}{\pa \over \pa t}\right)+{v\over 2L}\right]\int_{-1}^1d\mu \, \mu g
\eba
+{\a U\over 2v}\left[p{\pa \over \pa p}+2\right]\int_{-1}^1d\mu \, \mu g
+{\a \over 2}\left[p{\pa \over \pa p}+3\right]\int_{-1}^1d\mu \, \mu ^2g
\ebe
+{v(1-\mu ^2)\over 2L}{\partial g\over \partial \mu }=
{\pa \over \pa \mu }\left[D_{\mu \mu } {\pa g\over \pa \mu }\right]
\label{b7}
\ee
We note that Eqs. (\ref{b6}) and (\ref{b7}) are still exact.
\subsection{Diffusion approximation for cosmic shock waves}
With $|g|\ll F$ we approximate the anisotropy equation (\ref{b7}) to leading order by 

\bdm
\Gamma v\mu \left({\pa F\over \pa z}+{U\over c^2}{\pa F\over \pa t}\right)+\a \left[{1\over 3}-\mu ^2-{U\mu \over v}\right]p{\pa F\over \pa p}
\ebe
\simeq {\pa \over \pa \mu }\left[D_{\mu \mu } {\pa g\over \pa \mu }\right]
\label{c1}
\ee
In previous nonrelativistic ($\Gamma \simeq 1$) studies (Jokipii 1966, Hasselmann and Wibberenz 1968, Schlickeiser et al. 2007, Schlickeiser and Shalchi 2008) only the first term on the left-hand side of Eq. (\ref{c1}) has been considered, so that 

\be
v\mu {\pa F\over \pa z}\simeq {\pa \over \pa \mu }\left[D_{\mu \mu } {\pa g\over \pa \mu }\right]
\label{c2}
\ee
For gradual flows with non-zero values of $\a $ for an extended region of space, we have to consider also the term resulting from the momentum gradient $(\pa F/\pa p)$. This case will be considered elsewhere. 

Here we consider flows with non-zero values of $\a $ in a very limited region of space such as cosmic shock waves. We chose as laboratory frame the rest frame of the shock wave with the step-wise velocity profile 

\be
U(z)=
\begin{cases}  -U_1=\hbox{const.}  & \text{for $ 0< z\le \infty  $} \; \hbox{(upstream)}\\
-U_2=\hbox{const.} & \text{for $ -\infty \le z<0 $} \; \hbox{(downstream)}
\end{cases}
\label{c3}
\ee
with $U_2<U_1$. In this case the rate of adiabatic acceleration (\ref{a4}) 

\be
\a =\a _0\delta (z),\;\; \a _0=-(U_1\Gamma _1-U_2\Gamma _2)
\label{c4}
\ee
is non-zero only at the position of the shock. We therefore neglect in the anisotropy equation (\ref{c1}) the term resulting from the momentum gradient $(\pa F/\pa p)$, and additionally neglect the time derivate of $F$ as compared to the spatial gradient of $F$, i.e. 

\be
{U\over c^2}{\pa F\over \pa t}\ll {\pa F\over \pa z},
\label{c5}
\ee
so that Eq. (\ref{c1}) reduces to the streaming anisotropy contribution only 

\be
\Gamma v\mu {\pa F\over \pa z}\simeq {\pa \over \pa \mu }\left[D_{\mu \mu } {\pa g\over \pa \mu }\right],
\label{c6}
\ee
which differs from the nonrelativistic anisotropy equation (\ref{c2}) by the additional factor $\Gamma $. 

Integrating Eq. (\ref{c6}) provides  

\be
D_{\mu \mu }{\pa g\over \pa \mu }=c_0+{\Gamma v\mu ^2\over 2}{\pa F\over \pa z}
\label{c7}
\ee
The integration constant $c_0$ (with respect to $\mu $) is determined from the property that the left-hand side of this equation vanishes for $\mu =\pm 1$, yielding 
$c_0=-(\Gamma v/2)(\pa F/dz)$, providing 

\be
{\pa g\over \pa \mu }=-{\Gamma v(1-\mu ^2)\over 2D_{\mu \mu }(\mu )}{\pa F\over \pa z}
\label{c8}
\ee
A further integration over $\mu $ together with Eq. (\ref{b4}) provides for the \kr anisotropy 

\bdm
g(\XX, p,\mu ,t)\simeq {\Gamma v\over 4}\Bigl[\int_{-1}^1d\mu \, {(1-\mu )(1-\mu ^2)\over D_{\mu \mu }(\mu )}
\ebe
-2\int_{-1}^{\mu }dx \, {(1-x^2)\over D_{\mu \mu }(x)}\Bigr]{\pa F\over \pa z},
\label{c9}
\ee
which is determined by the spatial gradient of the isotropic distribution function $F(X,Y,z,p,t)$ with respect to $z$. 
\subsection{Anisotropy moments}
With the anisotropy (\ref{c9}) we readily determine the moments needed in the diffusion-convection transport equation (\ref{b6}) as 

\be
\int_{-1}^1d\mu \, \mu g=-{\Gamma vK_0\over 4}{\pa F\over \pa z}
\label{c10}
\ee
and 

\be
\int_{-1}^1d\mu \, \mu ^2g=-{\Gamma vK_1\over 6}{\pa F\over \pa z}
\label{c11}
\ee
in terms of the two ($n=0,1$) integrals 

\be
K_n=\int_{-1}^1d\mu \, {\mu ^n(1-\mu ^2)^2\over D_{\mu \mu }(\mu )}
\label{c12}
\ee
For the often considered case of symmetric pitch-angle \fo coefficients $D_{\mu \mu }(-\mu )=D_{\mu \mu }(\mu )$ the integral $K_1=0$ vanishes. 
\subsection{Diffusion-convection transport equation} 
For further reduction of the diffusion-convection transport equation (\ref{b6}) it is convenient to introduce as new momentum variable  

\be
s=\ln p
\label{d1}
\ee
and to define the convection and diffusion operators

\be
{\cal C}F=-{\a \over 3}p{\pa F\over \pa p}+{v\over 2L}\int_{-1}^1d\mu \,\mu g=-{\a \over 3}{\pa F\over \pa s}
+{v\over 2L}\int_{-1}^1d\mu \,\mu g
\label{d2}
\ee
and 

\bdm
{\cal D}F={v\Gamma \over 2}{\pa \over \pa z}\int_{-1}^1d\mu \, \mu g
-{\a U\over 2v}\left[p{\pa \over \pa p}+2\right]\int_{-1}^1d\mu \, \mu g
\eba
-{\a \over 2}\left[p{\pa \over \pa p}+3\right]\int_{-1}^1d\mu \, \mu ^2g
={v\Gamma \over 2}{\pa \over \pa z}\int_{-1}^1d\mu \, \mu g
\ebe
-{\a U\over 2v}\left[{\pa \over \pa s}+2\right]\int_{-1}^1d\mu \, \mu g
-{\a \over 2}\left[{\pa \over \pa s}+3\right]\int_{-1}^1d\mu \, \mu ^2g,
\label{d3}
\ee
respectively. With the assumption (\ref{c5}) we then can write the diffusion-convection transport equation (\ref{b6}) in the compact form 

\be
\Gamma \left[{\pa F\over \pa t}+U{\pa F\over \pa z}\right]+{\cal C}F+{\cal D}F+{\cal R}F=S(\XX ,p,t)
\label{d4}
\ee
Using for any momentum dependent quantity $M(s)$ the identity 

\bdm
{1\over v}{\pa \over \pa s}\left[A(s)M(p)\right]={\pa \over \pa s}\left[{A(s)\over v}M(p)\right]+{A(s)\over v\gamma ^2}M(p)
\ebe
={\pa \over \pa s}\left[{A(s)\over v}M(p)\right]+{A(s)(1-\beta ^2)\over v}M(p)
\label{d5}
\ee
with the \kr particle Lorentz factor $\gamma =(1-\beta ^2)^{-1/2}$ and the dimensionless particle velocity $\beta =v/c$, 
the diffusion operator (\ref{d3}) reads

\bdm
{\cal D}F={v\Gamma \over 2}{\pa \over \pa z}\int_{-1}^1d\mu \, \mu g
-{\a \over 2}{\pa \over \pa s}\left[{U\over v}\int_{-1}^1d\mu \, \mu g+\, \int_{-1}^1d\mu \, \mu ^2g\right]
\ebe
-{\a \over 2}\left[{U\over v}(3-\beta ^2)\int_{-1}^1d\mu \, \mu g+\, 3\int_{-1}^1d\mu \, \mu ^2g\right]
\label{d6}
\ee
With the moments (\ref{c8}) - (\ref{c9}) we derive after straightforward but tedious algebra 

\bdm
{\cal D}F=-\Gamma {\pa \over \pa z}\left[\Gamma \kappa _{zz}{\pa F\over \pa z}\right]
+\Gamma {\pa \over \pa s}\left[\kappa _{sz}{\pa F\over \pa z}\right]
\ebe
+{\Gamma \a v\over 4}\left(K_1+{3-\beta ^2\over 2}{U\over v}K_0\right){\pa F\over \pa z},
\label{d7}
\ee
where we introduce the two diffusion coefficients 

\bdm
\kappa _{zz}={v^2K_0\over 8},
\ebe
\kappa _{sz}=\kappa _{zs}={\a v\over 12}\left(K_1+{3U\over 2v}K_0\right)
\label{d8}
\ee
Likewise, the convection operator (\ref{d2}) becomes with the first moment (\ref{c10})

\be
{\cal C}F=-{\Gamma \kappa _{zz}\over L}{\pa F\over \pa z}-{\a \over 3}{\pa F\over \pa s}
\label{d9}
\ee
Inserting the operators (\ref{d7}) and (\ref{d9}) provides for the diffusion-convection transport equation (\ref{d4})

\bdm
\Gamma {\pa F\over \pa t}+\Gamma \left[U-{\kappa _{zz}\over L}+{\a v\over 4}\left(K_1+{3-\beta ^2\over 2}{U\over v}K_0\right)\right]
{\pa F\over \pa z}+{\cal R}F
\ebe
-{\a \over 3}{\pa F\over \pa s}-\Gamma {\pa \over \pa z}\left[\Gamma \kappa _{zz}{\pa F\over \pa z}\right]
+\Gamma {\pa \over \pa s}\left[\kappa _{zs}{\pa F\over \pa z}\right]=S(\XX ,p,t)
\label{d10}
\ee
Applying the identity 

\be
{\pa \over \pa s}\left[\kappa M\right]={1\over p^2}{\pa \over \pa p}\left[p^3\kappa M\right]-3\kappa M,
\label{d11}
\ee
for any momentum dependent quantity $M(s)$ to the last term on the left-hand side of Eq. (\ref{d10}) we obtain 

\bdm
{\pa \over \pa s}\left[\kappa _{sz}{\pa F\over \pa z}\right]=
{1\over p^2}{\pa \over \pa p}p^3\left[\kappa _{sz}{\pa F\over \pa z}\right]-3\kappa _{sz}{\pa F\over \pa z}
\ebe
={1\over p^2}{\pa \over \pa p}p^2\left[\kappa _{sz}p{\pa F\over \pa z}\right]
-3\kappa _{sz}{\pa F\over \pa z}
\label{d12}
\ee
Consequently, we find as alternative form for the diffusion-convection transport equation (\ref{d10})

\bdm
\Gamma {\pa F\over \pa t}+{\cal R}F-\Gamma {\pa \over \pa z}\left[\Gamma \kappa _{zz}{\pa F\over \pa z}\right]
+{\Gamma \over p^2}{\pa \over \pa p}p^2\kappa _{sz}p{\pa F\over \pa z}
\eba
+\Gamma \left[U-{\kappa _{zz}\over L}-3\kappa _{sz}+{\a v\over 4}\left(K_1+{3-\beta ^2\over 2}{U\over v}K_0\right)\right]
{\pa F\over \pa z}
\ebe
-{\a \over 3}{\pa F\over \pa s}=S(\XX ,p,t)
\label{d13}
\ee
With the diffusion coefficients (\ref{d8}) we obtain for the difference 

\bdm
{\a v\over 4}\left(K_1+{3-\beta ^2\over 2}{U\over v}K_0\right)-3\kappa _{sz}=-{\a UK_0\beta ^2\over 8}
\ebe
=-{\a U\over c^2}\kappa _{zz}=-\kappa _{zz}{d\Gamma \over dz},
\label{d14}
\ee
where we used Eq. (\ref{a4}). The diffusion-convection transport equation (\ref{d13}) then becomes 

\bdm
\Gamma {\pa F\over \pa t}-{\kappa _{zz}\Gamma \over L}{\pa F\over \pa z}+\Gamma \left[U{\pa F\over \pa z}-{\a p\over 3\Gamma }{\pa F\over \pa p}\right]
-{d\Gamma \over dz}\Gamma \kappa _{zz}{\pa F\over \pa z}
\ebe
-\Gamma {\pa \over \pa z}\left[\Gamma \kappa _{zz}{\pa F\over \pa z}\right]+{\Gamma \over p^2}{\pa \over \pa p}p^2\kappa _{pz}{\pa F\over \pa z}
+{\cal R}F=S(\XX ,p,t)
\label{d15}
\ee
with 

\bdm
\kappa _{pz}=p\kappa _{sz}=\a K_{pz},
\ebe
K_{pz}={vp\over 12}\left(K_1+{3U\over 2v}K_0\right)
\label{d16}
\ee
We note the identity 

\bdm
U{\pa F\over \pa z}-{\a p\over 3\Gamma }{\pa F\over \pa p}={1\over \Gamma }{\pa \over \pa z}\left[\Gamma UF\right]-{F\over \Gamma }{d\over dz}\left(\Gamma U\right)
\ebe
-{1\over \Gamma p^2}{\pa \over \pa p}\left({\a p^3F\over 3}\right)+{\a F\over \Gamma }=
{1\over \Gamma }{\pa \over \pa z}\left[\Gamma UF\right]-{1\over \Gamma p^2}{\pa \over \pa p}\left({\a p^3F\over 3}\right),
\label{d17}
\ee
because 

\be
{d\over dz}\left(\Gamma U\right)=\Gamma {dU\over dz}+U{d\Gamma \over dz}=\Gamma {dU\over dz}\left[1+{U^2\over c^2}\Gamma ^2\right]=\a 
\label{d18}
\ee
Likewise,

\bdm
{d\Gamma \over dz}\kappa _{zz}\Gamma {\pa F\over \pa z}+\Gamma {\pa \over \pa z}\left[\Gamma \kappa _{zz}{\pa F\over \pa z}\right]
\ebe
=\left({d\Gamma \over dz}+\Gamma {\pa \over \pa z}\right)\left[\Gamma \kappa _{zz}{\pa F\over \pa z}\right]={\pa \over \pa z}\left[\Gamma ^2\kappa _{zz}{\pa F\over \pa z}\right]
\label{d19}
\ee
With these two identities the diffusion-convection transport equation (\ref{d15}) finally reads

\bdm
\Gamma {\pa F\over \pa t}-{\kappa _{zz}\Gamma \over L}{\pa F\over \pa z}+{\pa \over \pa z}\left[\Gamma \left(UF-\Gamma \kappa _{zz}{\pa F\over \pa z}\right)\right]
\ebe
+{1\over p^2}{\pa \over \pa p}\left[p^2\kappa _{pz}\Gamma {\pa F\over \pa z}-{\a p^3F\over 3}\right]
+{\cal R}F=S(\XX ,p,t)
\label{d20}
\ee
Eq. (\ref{d20}) is the first important new result of this study: the diffusion-convection transport equation of \krs in aligned parallel flows of arbitrary speed containing magnetostatic slab turbulence with the \kr phase space coordinates taken in the mixed comoving coordinate system. It is particularly appropriate to investigate \kr particle acceleration in parallel relativistic flows. 

In the limit of nonrelativistic flows $U(z)\ll c$ so that $\Gamma \simeq 1$. the transport equation (\ref{d20}) becomes 

\bdm
{\pa F\over \pa t}-{\kappa _{zz}\over L}{\pa F\over \pa z}+{\pa \over \pa z}\left[UF-\kappa _{zz}{\pa F\over \pa z}\right]
\ebe
+{1\over p^2}{\pa \over \pa p}\left[p^2\kappa _{pz}{\pa F\over \pa z}-{\a p^3F\over 3}\right]
+{\cal R}F=S(\XX ,p,t), 
\label{d21}
\ee
which differs from the transport theory used in earlier nonrelativistic diffusive shock acceleration theory (Axford et al. 1977, Krymsky 1987, Blandford and Ostriker 1978, Bell 1978, Drury 1983) by the additional third last term on the left-hand side involving $\kappa _{pz}$, which results from our correct handling of the connection coefficients in Eq. (\ref{a3}). As we will demonstrate below, this additional term  provides a major modification of the resulting differential momentum spectrum of accelerated particles in the nonrelativistic flow limit at nonrelativistic particles momenta: instead of a power law distribution of accelerated particles at the shock a Lorentzian distribution function results, which at large momenta then approaches the power law distribution inferred in earlier acceleration theories for nonrelativistic shock speeds.   
\subsection{Slab Alven waves}
There are four different magnetostatic slab Alven waves: forward and backward propagating waves which each can be 
left- or right-handed circularly polarized, respectively. In terms of the cross helicity $H_c$ and the magnetic helicities 
$\sigma _{\pm }$ of forward and backward moving Alfven waves, and power-law type wave intensities $I\propto \kpa ^{-s}$ with 
$s\in (1,2)$ with the same spectral index $s$ of all four waves (isospectral turbulence), 
the two integrals (\ref{c12}) are given by (Dung and Schlickeiser 1990, Schlickeiser 2002) 

\be
K_n={64\over s-1}{(R_Lk_{\rm min})^{2-s}\over vk_{\rm min}}\left({B_0\over \delta B}\right)^2S_n(H_c,\sigma _+,\sigma _-)
\label{e1}
\ee
for $R_Lk_{\rm min}\le 1$ with 

\bdm
S_0={2\over (2-s)(4-s)G(H_c,\sigma _+,\sigma _-)}
\ebe
S_1={Z\left[\sigma _++\sigma _-+H_c(\sigma _+-\sigma _-)\right]\over (3-s)(5-s)G(H_c,\sigma _+,\sigma _-)},
\label{e2}
\ee
where 

\be
G(H_c,\sigma _+,\sigma _-)=4-\left[\sigma _++\sigma _-+H_c(\sigma _+-\sigma _-)\right]^2
\label{e3}
\ee
is always positive. $R_L=v/|\Omega |=pc/(|q|B_0)$ is the cosmic ray gyroradius, $Z=q/|q|$ the sign of the cosmic ray particle and 
$k_{\rm min}$ the smallest wavenumber of the Alfven waves with total magnetic field strength $(\delta B)^2$. It is obvious that the two integrals (\ref{e1}) exhibit the same momentum dependence $K_n\propto (p/|q|)^{2-s}v^{-1}$. Moreover, $S_1/Z$ is positive for 
$H_c>(\sigma _++\sigma _-)/(\sigma _--\sigma _+)$ and negative for $H_c<(\sigma _++\sigma _-)/(\sigma _--\sigma _+)$. 

Defining the maximum cosmic ray momentum with $q=Qe$

\be
p_m={|Q|eB_0\over k_{\rm min}c}=1.5\cdot 10^{16}|Q|B(\mu \hbox{G})\lambda _{100}\;\; {\hbox{eV}\over c},
\label{e4}
\ee
where $\lambda _{\rm max}=2\pi k^{-1}_{\rm min}=100\lambda _{100}$ pc denotes the maximum wavelength of the Alfven waves, we 
obtain for the integrals (\ref{e1}) for $p\le p_m$ 

\be
K_n=K{S_n\over v}\left({p\over p_m}\right)^{2-s}
\label{e5}
\ee
with the constant length

\be
K={32\over \pi (s-1)}\left({B_0\over \delta B}\right)^2\lambda _{\rm max}=3.1\cdot 10^{21}\lambda _{100}\left({B_0\over \delta B}\right)^2
\label{e6}
\ee
For larger momenta $p>p_m$ the integrals (\ref{e1}) are zero. 

Consequently, we find for the diffusion coefficients (\ref{d8}) and (\ref{d16})

\bdm
\kappa _{zz}={vK\over 8}S_0\left({p\over p_m}\right)^{2-s},
\ebe
\kappa _{pz}=\a K_{pz},\;\; K_{pz}={pK\over 12}\left(S_1+{3US_0\over 2v}\right)\left({p\over p_m}\right)^{2-s}
\label{e7}
\ee
\section{Particle acceleration at relativistic shock waves}
We adopt the step-like shock profile (\ref{c3}) and assume particle injection $S(\XX ,p,t)=S_0\delta (z)$ at the position of the shock only with the injection momentum spectrum $S(p)$. Moreover, we assume spatially constant flow velocities and diffusion coefficients in the upstream and downstream region. 

In the steady-state case with no losses (${\cal RF}=0$) and a uniform background magnetic field ($L=\infty $) the diffusion-convection transport equation (\ref{d20}) in the rest frame of the shock wave reduces to 

\bdm
{\pa \over \pa z}\left[\Gamma \left(UF-\Gamma \kappa _{zz}{\pa F\over \pa z}\right)\right]
-{\a _0\delta (z)\over p^2}{\pa \over \pa p}\left[p^2K_{pz}\Gamma {\pa F\over \pa z}-{p^3F\over 3}\right]
\ebe
=S(p)\delta (z=0)
\label{f1}
\ee
We will solve Eq. (\ref{f1}) by the same method as for nonrelativistic parallel step-like shock waves (Axford et al. 1977, Krymsky 1987, Blandford and Ostriker 1978, Bell 1978, Drury 1983). 

For the upstream region $z>0$ the steady-state transport equation (\ref{f1}) yields 

\be
F_1(z,p)=F_0(p)\exp \left[-{U_1z\over \Gamma _1\kappa _{zz,1}}\right],
\label{f2} 
\ee
which approaches zero far upstream $z\to \infty $. 

For the downstream region $z<0$ the solution of the steady-state transport equation (\ref{f1}) is given by 

\be
F_2(z,p)=F_0(p),
\label{f3} 
\ee
which is finite far downstream $z\to -\infty $. At the position of the shock

\be 
F_1(z=0,p)=F_2(z=0,p)=F_0(p)
\label{f300}
\ee
the distribution function is continuous. 
\subsection{Momentum spectrum of accelerated particles at the shock}
The particle momentum spectrum $F_0(p)$ at the position of the shock is obtained by integrating the transport 
equation (\ref{f1}) from $z=-\eta $ to $z=\eta $ and considering the limit $\eta \to 0$. This provides the continuity condition for the \kr streaming density at the shock 

\bdm 
\Gamma _1\left(U_1F_1-\Gamma _1\kappa _{zz,1}{\pa F_1\over \pa z}\right)_{0^+}-\Gamma _2\left(U_2F_2-\Gamma _2\kappa _{zz,2}{\pa F_2\over \pa z}\right)_{0^-}
\ebe
-{\a _0\over p^2}{\pa \over \pa p}\left[p^2K_{pz,1}\Gamma _1\left({\pa F_1\over \pa z}\right)_{0^+}-{p^3F_0\over 3}\right]=S(p)
\label{f4} 
\ee
With the up- and downstream solutions (\ref{f2}) and (\ref{f3}) we obtain 

\bdm
\Gamma _2U_2F_0(p)+{U_1\Gamma _1-U_2\Gamma _2\over 3p^2}{d\over dp}\left(p^3+{3p^2K_{pz,1}U_1\over \kappa _{zz,1}}\right)F_0(p)
\ebe
=S(p)
\label{f5}
\ee
where we inserted $\a _0$ from Eq. (\ref{c4}). 

According to Eqs. (\ref{e7}) we find for the case of slab Alfven waves discussed in Sect. 3.4 for the ratio 

\bdm
{3K_{pz,1}U_1\over \kappa _{zz,1}}=p{U_1\over v}\left[{2S_1\over S_0}+{3U_1\over v}\right]
\ebe
=p{U_1\over v}\left[ZR(s,\sigma _+,\sigma _-,H_c)+{3U_1\over v}\right]
\label{f6}
\ee
with the helicity dependent function  

\be
R(s,\sigma _+,\sigma _-,H_c)={(2-s)(4-s)\over (3-s)(5-s)}\left[\sigma _++\sigma _-+H_c(\sigma _+-\sigma _-)\right]
\label{f7}
\ee
The value of the bracket of this function is not greater than $2$ for all helicity values, so the function is limited to values 

\be
|R_{\rm max}|\le {2(2-s)(4-s)\over (3-s)(5-s)},
\label{f8}
\ee
which for all values of $s\in [1,2)$ is smaller than $0.75$. 

In the case of symmetric pitch-angle \fo coefficients $D_{\mu \mu }(-\mu )=D_{\mu \mu }(\mu )$ so that $K_1=0$ the ratio (\ref{f6}) simplifies to 

\be
{3K_{pz,1}U_1\over \kappa _{zz,1}}=3p{U_1^2\over v^2},
\label{f9}
\ee
which agrees with the limit for $R=0$ of Eq. (\ref{f6}). 

For both cases Eq. (\ref{f5}) then reads 

\be
F_0(p)+{1\over \psi p^2}{d\over dp}\left(p^2T(p)F_0(p)\right)={S(p)\over \Gamma _2U_2}
\label{f10}
\ee
where we define the ratio 

\be
\psi ={3U_2\Gamma _2\over U_1\Gamma _1-U_2\Gamma _2}
\label{f11}
\ee
and the function 

\bdm
T(p)=p\left(1+{\beta_1\over \beta }[ZR+{3\beta_1\over \beta}]\right)=p\Bigl(1+
\ebe
{mc\beta_1\sqrt{1+(p/mc)^2}\over p}\left[ZR+{3mc\beta _1\sqrt{1+(p/mc)^2}\over p}\right]\Bigr)
\label{f12}
\ee
with $\beta _1=U_1/c$ and $\beta =v/c$. Setting 

\be
M(p)=p^2T(p)F_0(p)
\label{f13}
\ee
provides for Eq. (\ref{f10}) 

\be
T(p){dM(p)\over dp}+\psi M(p)={\psi p^2S(p)T(p)\over \Gamma _2U_2}
\label{f14}
\ee
Eq. (\ref{f14}) suggests to introduce as momentum variable 

\bdm
w(p)=\int_0^p{dp^{'}\over T(p^{'})}
\eba
=\int_{mc\over p}^\infty {dy\over y\left[1+\beta _1\sqrt{1+y^2}[ZR+3\beta _1\sqrt{1+y^2}]\right]}
\ebe
=\int_0^{\beta } dt\, {t\over (1-t^2)\left[t^2+\beta _1ZRt+3\beta ^2_1]\right]},
\label{f15}
\ee
implying 

\be
{dw\over dp}={1\over T(p)}
\label{f16}
\ee
Eq. (\ref{f14}) then becomes 

\be
{d\over dw}\left[M(w)e^{\psi w}\right]=Q(w)
\label{f17}
\ee
with the source term 

\be
Q(w)={3p^2(w)S(p(w))T(p(w))e^{\psi w}\over U_1\Gamma _1-U_2\Gamma _2}
\label{f18}
\ee
Eq. (\ref{f18}) is solved for any injection spectrum $S(p)$ by 

\be
M(w)=\int_0^\infty dw^{'}\, G(w,w^{'})Q(w^{'})
\label{f19}
\ee
in terms of the Green$^{'}$s function $G(w,w^{'})$ obeying 

\be
{d\over dw}\left[G(w,w^{'})e^{\psi w}\right]=\delta (w-w^{'})
\label{f20}
\ee
The solution of Eq. (\ref{f20}) is given by 

\be
G(w,w^{'})=H\left[w-w^{'}\right]e^{-\psi w},
\label{f21}
\ee
where $H[x]$ denotes Heaviside's step function. With Eqs. (\ref{f21}) and (\ref{f18}) we obtain for Eq. (\ref{f19}) 

\bdm
M(w)=e^{-\psi w}\int_0^wdw^{'}\, Q(w^{'})={3\over U_1\Gamma _1-U_2\Gamma _2}
\ebe
\times \int_{p^{'}}^pdt\, t^2S(t))\exp \left[-\psi \int_{t}^{p}{dx\over T(x)}\right],
\label{f22}
\ee
so that according to Eq. (\ref{f13}) 

\be
F_0(p)={3\over U_1\Gamma _1-U_2\Gamma _2}{1\over p^2T(p)}
\int_{p^{'}}^p dt \, t^2S(t))e^{-\psi \int_{t}^{p}{dx\over T(x)}}
\label{f23}
\ee
\subsection{Monomentum injection}
For a monomomentum injection spectrum 

\be
S(p)=S_0\delta (p-p_0),
\label{f24}
\ee
the solution (\ref{f23}) becomes  

\be
F_0(p\ge p_0)={3S_0\over U_1\Gamma _1-U_2\Gamma _2}{p_0^2\over p^2T(p)}e^{-\psi I(p,p_0)}
\label{f25}
\ee
with the integral 

\bdm
I(p,p_0)=\int_{p_0}^{p}{dx\over T(x)}
\eba
=\int_{mc\over p}^{mc\over p_0}{dy\over y\left[1+\beta _1\sqrt{1+y^2}[ZR+3\beta _1\sqrt{1+y^2}]\right]}
\ebe
=\int_{\beta _0}^{\beta } dt\, {t\over (1-t^2)\left[t^2+\beta _1ZRt+3\beta ^2_1]\right]},
\label{f26}
\ee
The integral (\ref{f26}) can be solved in closed form (see Appendix A), but asymptotic expressions for relativistic and nonrelativistic 
\kr particle momenta can be directly deduced from approximating the integrand in Eq.(\ref{f26}). 

For the differential number density of accelerated particles $N(p)=4\pi p^2F(p)$ the solution (\ref{f26}) implies 

\bdm
N_0(p\ge p_0)=4\pi p^2F(p\ge p_0)
\ebe
={4\pi S_0\psi \over U_2\Gamma _2}{p_0^2\over T(p)}e^{-\psi I(p,p_0)}
\label{f27}
\ee
at the position of the shock $z=0$, and the up- and down-stream number densities 

\be
N_1(z>0,p)=N_0(p\ge p_0)\exp \left[-{U_1z\over \Gamma _1\kappa _{zz,1}}\right]
\label{f28} 
\ee
and

\be
N_2(z<0,p)=N_0(p\ge p_0),
\label{f29} 
\ee
where the ratio (\ref{f11}) is given by 

\be
\psi ={3\over \sqrt{r^2-\beta _1^2\over 1-\beta _1^2}-1}
\label{f30} 
\ee
in terms of shock wave compression ratio $r=U_1/U_2=\beta _1/\beta _2$.

In the following sections we will investigate some interesting limits of this solution. 
\section{Symmetric pitch-angle Fokker-Planck coefficients ($R=0$)}
For symmetric pitch-angle \fo coefficients ($R=0$) the integral (\ref{f26}) reduces to 

\bdm
I(R=0)=\int_{p_0/mc}^{p/mc}{dy\over y\left[1+3\beta _1(1+{1\over y^2})\right]}
\ebe
={1\over 2}\int_{p_0^2\over m^2c^2}^{p^2\over m^2c^2}{ds\over 3\beta _1+(1+3\beta _1^2)s}
={1\over 2(1+3\beta _1^2)}\ln {1+{p^2\over p_c^2}\over 1+{p_0^2\over p_c^2}}
\label{k1}
\ee
with the characteristic momentum 

\be
p_c(\beta _1)=\sqrt{3\beta _1^2\over 1+3\beta _1^2}mc
\simeq \begin{cases}  \sqrt{3}mU_1 & \text{for $ \beta _1\ll 1 $} \\
mc & \text{for $ \Gamma _1\gg 1 $}
\end{cases}
\label{k5}
\ee
Then the differential number density at the shock (\ref{f27}) becomes the Lorentzian-type distribution function 

\be
N_0(p\ge p_0)=A_0p\left[1+({p\over p_c})^2\right]^{-\rho }
\label{k2}
\ee
with 

\be
A_0={4\pi S_0\psi p_0^2\over U_2\Gamma _2(1+3\beta _1^2)p_c^2}\left[1+{p_0^2\over p_c^2}\right]^{\rho -1},
\label{k3}
\ee

\be
\rho={\psi \over 2(1+3\beta _1^2)}+1
\label{k4}
\ee

\begin{figure}[htbp]
\centering
\includegraphics[width=0.5\textwidth]{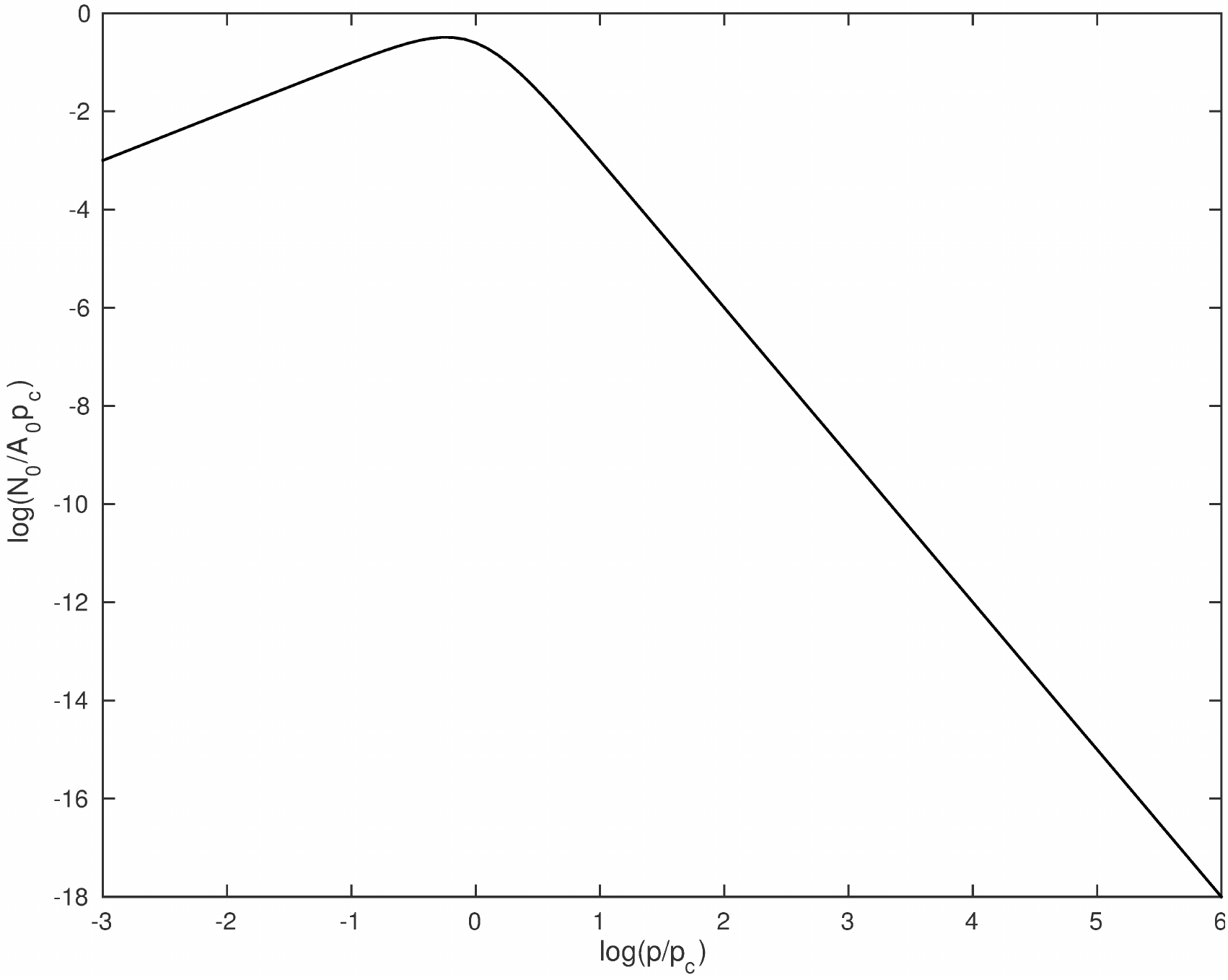}
\caption{Differential number density (\ref{k2}) of accelerated particles at the shock as a function of $p/p_c$ in the case $R=0$ 
for the adopted spectral index value $\rho =2$ and injection momentum $p_0/p_c=10^{-3}$.}
\end{figure}

In Fig. 1 we illustrate the Lorentzian differential number density of accelerated particles at the shock as a function of $p/p_c$. 
For particle momenta $p_0\le p\le p_c$ the Lorentzian distribution (\ref{k2}) increases linearly with momentum, $N_0(p_0\le p\le p_c)\simeq A_0p$, whereas for large momenta $p\ge p_c$ it approaches the decreasing power law distribution 

\be
N_0(p\ge p_c)\simeq A_0p_c\left({p\over p_c}\right)^{-\xi  }
\label{k6}
\ee
with 

\bdm
\xi =2\rho-1=1+{\psi \over 1+3\beta _1^2}
\ebe
=1+ {3\over (\Gamma _1\sqrt{r^2-\beta _1^2}-1)(1+3\beta _1^2)}
\label{k61}
\ee
Notice that for relativistic ($\Gamma _1\gg 1$) shock speeds the characteristic momentum (\ref{k5}) coincides with $mc$, so that in this case 
the no decreasing power law distributions for nonrelativistic shock accelerated particles result. 

The Lorentzian distribution (\ref{k2}) attains its maximum value 

\be
N_{0,\rm max}=A_0p_c{(2\rho -1)^{\rho -{1\over 2}}\over (2\rho )^\rho }
\label{k7}
\ee
at $p_{\rm max}=p_c/\sqrt{2\rho -1}$. The total number of accelerated particles at the shock is given by 

\bdm
N_{0,\rm tot}=\int_{p_0}^\infty N_0(p\ge p_0)={A_0p_c^2\over 2(\rho -1)}\left[1+({p_0\over p_c})^2\right]^{1-\rho }
\ebe
={4\pi S_0p_0^2\over U_2\Gamma _2}={4\pi S_0p_0^2\sqrt{r^2-\beta _1^2}\over c\beta _1}
\label{k8}
\ee
The injection rate of charged particles in a partially ionized (with ionization fraction $\tau _i=n_{e1}/n_{\rm tot 1}$) cold upstream medium is related to the upstream particle number flux $n_{\rm tot 1}U_1$ as 

\be
S_0\delta (p-p_0)={\tau _i\tau _en_{\rm tot 1}U_1\over p_1^2}\delta (p-p_1)={\tau _i\tau _en_{\rm tot 1}c\beta _1\over p_1^2}\delta (p-p_1),
\label{k9}
\ee
where the injection efficiency $\tau _e\ll 1$ indicates the small fraction of ionized upstream particles being accelerated. 
Then the total number of accelerated particles at the shock (\ref{k8}) is 

\be
N_{0,\rm tot}=4\pi \tau _i\tau _In_{\rm tot 1}\sqrt{r^2-\beta _1^2}
\label{k10}
\ee
\subsection{Nonrelativistic shock waves}
For nonrelativistic shock velocities $\beta _1\ll 1$, so that $\Gamma _{1}\simeq 1$, Eq. (\ref{k4}) becomes 

\be
\rho \simeq {\psi _0\over 2}+1={2r+1\over 2(r-1)},
\label{k11}
\ee
where we used 

\be
\psi _0={3\over r-1}
\label{k12}
\ee
for the limit of Eq. (\ref{f30}) for $\beta _1\ll 1$. In this case the Lorentzian distribution function (\ref{k2}) reads 

\be
N_0(p\ge p_0)=A_0p\left[1+({p\over \sqrt{3}\beta _1mc})^2\right]^{-{2r+1\over 2(r-1)}}
\label{k13}
\ee
At momenta greater than the nonrelativistic characteristic momentum $p_c^{\rm nr}=\sqrt{3}\beta _1mc$ this function approaches the decreasing power law distribution 

\be
N_0(p\ge p_c^{\rm nr})\simeq A_0p_c^{\rm nr}\left({p\over p_c}\right)^{-\xi _0}
\label{k14}
\ee
with the spectral index

\be
\xi _0={r+2\over r-1}
\label{k15}
\ee
This spectral index agrees with the standard result for nonrelativistic shocks providing $\xi _0\ge 2$ for shocks in adiabatic electron-proton media with compression ratios $r\le 4$ and $\xi _0\ge 3/2$ for shocks in adiabatic electron-positron media with compression ratios $r\le 7$. 
\subsection{Relativistic shock waves}
For relativistic shock velocities with $\beta _1\simeq 1$ and  $\Gamma _{1}\gg 1$, Eq. (\ref{k4}) becomes 

\be
\rho \simeq 1+{\psi \over 8}\simeq 1+{3\over 8(\Gamma _1\sqrt{r^2-1}-1)}, 
\label{m1}
\ee
but now we have to distinguish between particle injection at nonrelativistic ($p_0\ll mc$) and at relativistic ($p_0\gg mc$) momenta. 

In the first case the Lorentzian distribution function (\ref{k2}) reads 

\be
N_0(p\ge p_0)=A_0p\left[1+({p\over mc})^2\right]^{-1-{3\over 8(\Gamma _1\sqrt{r^2-1}-1)}}
\label{m2}
\ee
At nonrelativistic particle momenta this function increases linearly in momentum. At relativistic particle momenta it approaches the decreasing power law distribution 

\be
N_0(p\ge mc)\simeq A_0mc\left({p\over mc}\right)^{-\xi (\Gamma _1\gg 1)}
\label{m3}
\ee
with the spectral index 

\be
\xi (\Gamma _1\gg 1)=1+{3\over 4(\Gamma _1\sqrt{r^2-1}-1},
\label{m4}
\ee
which for $\Gamma _1\gg 1$ approaches unity. We defer the discussion of this spectral index to the next section where the acceleration of relativistic \krs is investigated in more detail. 

If \krs are injected at relativistic momenta $p_0\gg mc$ the power law limit (\ref{k6}) of the distribution function (\ref{k2}) holds, so that  
with $p_c\simeq mc$ again Eq. (\ref{m3}) results. As $p_0\gg p_c$ the quantity (\ref{k3}) becomes 

\be
A_0={4\pi S_0\psi \over U_2\Gamma _2(1+3\beta _1^2)}\left[{p_0\over mc}\right]^{2\rho },
\label{m5}
\ee
providing for the distribution (\ref{m3})

\be
N_0(p\ge p_0\gg mc)\simeq {4\pi S_0\psi p_0\over U_2\Gamma _2(1+3\beta _1^2)}
\left({p\over p_0}\right)^{-\xi (\Gamma _1\gg 1)}
\label{m6}
\ee
\section{Relativistic cosmic rays}
For relativistic particle momenta the integral (\ref{f26}) can be solved for general values of the helicity dependent function $R$. 
We assume here that \kr particles are injected at relativistic momenta $p_0\gg mc$ with the monomomentum injection spectrum 
$S(p)=S_0\delta (p-p_0)$. With $p\ge p_0\gg mc$ the integral (\ref{f26}) reduces to 

\be
I(p\ge p_0\gg mc)\simeq {1\over 1+\beta _1ZR+3\beta _1^2}\ln \left({p\over p_0}\right)
\label{h1}
\ee
Consequently, the solution (\ref{f27}) becomes the power law distribution 

\be
N_0(p\ge p_0\gg mc)\simeq {4\pi S_0p_0\psi \over U_2\Gamma _2(1+\beta _1ZR+3\beta _1^2)}\left({p\over p_0}\right)^{-\xi }
\label{h2}
\ee
with the power law spectral index 

\bdm
\xi =1+{\psi \over 1+\beta _1ZR+3\beta _1^2}
\ebe
=1+{3\over \left(\Gamma _1\sqrt{r^2-\beta _1^2}-1\right)\left(1+\beta _1ZR+3\beta _1^2\right)}
\label{h3}
\ee
We first note that for $R=0$ the power law solution (\ref{h2}) agrees with the earlier derived expression (\ref{m6}) and that the spectral indices (\ref{h3}) and (\ref{k61}) agree.
\subsection{Nonrelativistic shock waves}
For nonrelativistic shock velocities $U_{1,2}\ll c$, so that $\Gamma _{1}\simeq 1+(\beta _1^2/2)$ and 
$\Gamma _{2}\simeq 1+(\beta _1^2/2r^2)$, the particle power law spectral index (\ref{h3}) to first order in $\beta _1\ll 1$ reduces to 

\be
\xi =1+{3[1-ZR\beta _1]\over r-1}={r+2-3ZR\beta _1\over r-1}
\label{h4}
\ee
To lowest order in $\beta _1$ we again reproduce the standard result for nonrelativistic shocks 

\be
\xi (\beta _1=0)={r+2\over r-1},
\label{h5}
\ee
providing $\xi \ge 2$ for shocks in adiabatic electron-proton media with compression ratios $r\le 4$ and $\xi \ge 3/2$ for shocks in adiabatic electron-positron media with compression ratios $r\le 7$. 

However, our result (\ref{h4}) gives a small (for turbulence spectral indices $s<2$) correction to this standard spectral index which is different for positively ($Z=1$) and negatively ($Z=-1$) charged \kr particles. Depending on the sign of helicity dependent function $R$ defined Eq. (\ref{f6}) this implies either a smaller or greater spectral index compared to the standard result (\ref{h5}). With the maximum value (\ref{f8}) the correction is at most  

\be
|\Delta \xi |\le{3R_{\rm max}\beta _1\over r-1}={6(2-s)(4-s)\over (3-s)(5-s)}{\beta _1\over r-1},
\label{h6}
\ee
which for a Kolmogorov turbulence spectral index $s=5/3$ gives 

\be
|\Delta \xi |\le{3R_{\rm max}\beta _1\over r-1}=1.05{\beta _1\over r-1},
\label{h7}
\ee
For most adiabatic shocks with $r>1.1$ this is negligibly small as $|\Delta \xi |\le 10.5\beta _1$.  
\subsection{Relativistic shock waves}
The determination of the power law spectral indices (\ref{h3}) and (\ref{k61}) require the knowledge of the shock compression ratio $r=\beta _1/\beta _2$ which for relativistic shocks depends for any given shock speed $\beta _1$ in a non-trivial way on the equations of state of the up- and downstream fluids as shown for hyrodynamical shocks by Peacock (1981), Heavens and Drury (1988) and Kirk and Duffy (1999). The jump conditions for relativistic magnetohydrodynamic shocks in gyrotropic plasmas were studied by Double et al. (2004) and Gerbig and Schlickeiser (2011), including the pressure anisotropy $\chi =P_{\perp }/P_{\parallel }$ of the upstream and downstream gas pressures adopting adiabatic equation of states of the up- and down-stream gas with adiabatic indices $\kappa _{1,2}$. For a parallel ultrarelativistic shock ($\Gamma _1\gg 1$, $\beta _1\simeq 1$) Gerbig and Schlickeiser (2011) found the relations between the downstream parallel plasma beta $\beta _{\parallel 2}=8\pi P_{\parallel 2}/B_0^2$, the 
compression ratio $r=\beta _1/\beta _2$ and the anisotropies $\chi _{1,2}$. 

For illustrating our results we consider here only the case of an ultrarelativistic shock $\Gamma _1\gg 1$ and a relativistic downstream medium with adiabatic index $4/3$, so that $\beta _1=3\beta _2$ (Blandford and McKee 1976) or $r=3$. In this case we obtain for the power law spectral indices (\ref{h3}) and (\ref{k61}) 

\bdm
\xi (\Gamma _1\gg 1)=1
\eba
+{3\over \left[\sqrt{8\Gamma _1^2+1}-1\right]\left[4-{3\over \Gamma _1^2}+ZR\sqrt{1-{1\over \Gamma _1^2}}\right]}
\ebe
\simeq 
1+{3\over 2\sqrt{2}[4+ZR]\Gamma _1}
\label{i1}
\ee

\begin{figure}[htbp]
\centering
\includegraphics[width=0.5\textwidth]{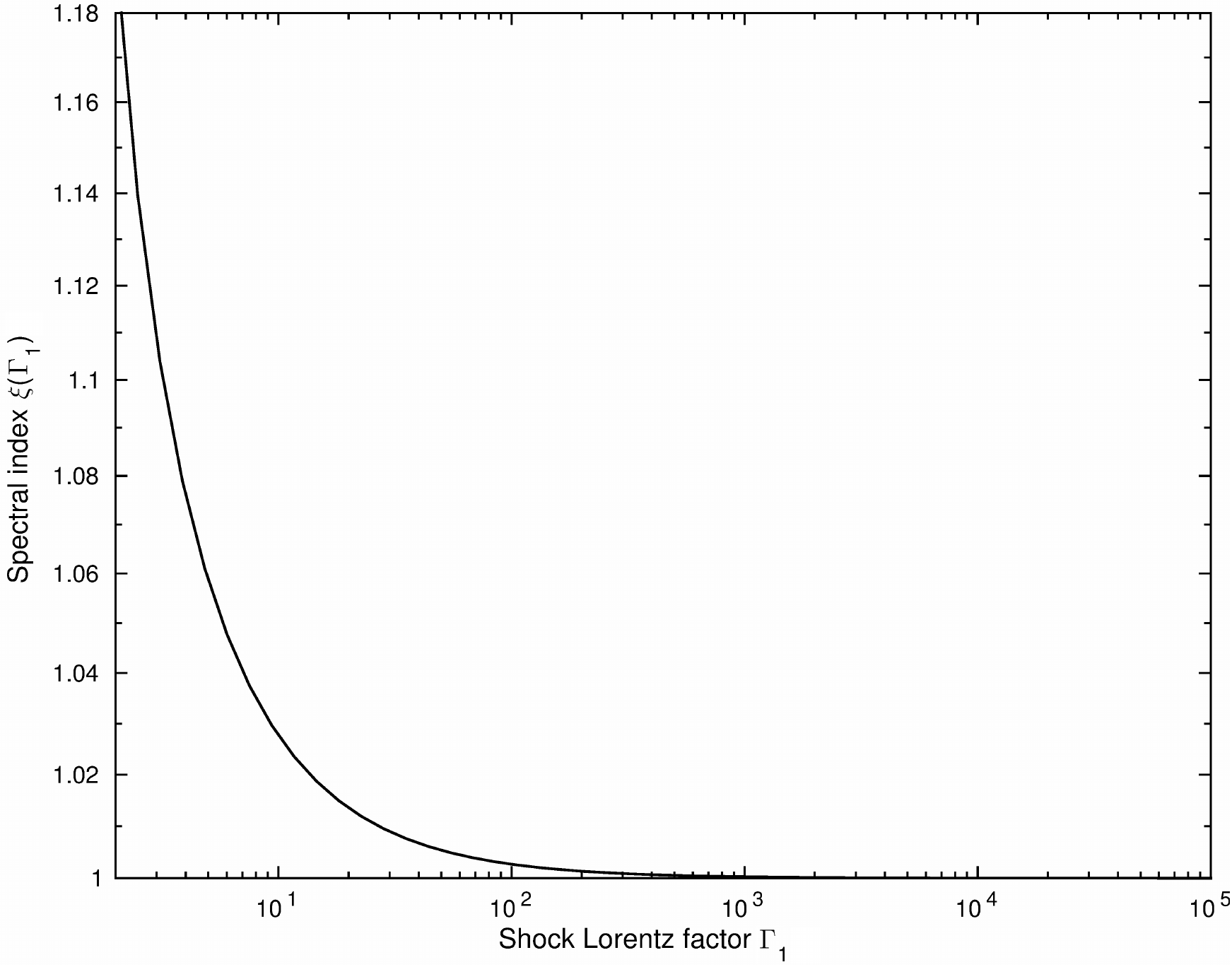}
\caption{Power law spectral index (\ref{i1}) of relativistic particles accelerated at an ultrarelativistic shock for the case $R=0$ as a function of the shock Lorentz factor $\Gamma _1$.}
\end{figure}

In Fig. 2 we calculate this spectral index for the case $R=0$ for relativistic shocks with $\Gamma _1\ge 2$, indicating spectral index values close to unity. Possible modifications and charge-sign dependencies (i.e. the case $R\ne 0$) will be considered elsewhere. Due to the dominating $\Gamma ^{-1}$ dependence of $\xi -1$ the limit $\xi \simeq 1$ is reached for $\Gamma _1>10$. 

Our result of flat spectral indices with $\xi \simeq 1$ for ultrarelativistic shock disagrees strongly with the earlier established universal spectral index value $\xi \in [2.25-2.30]$ from the eigenfunction and Monte Carlo simulation studies (for review see Kirk and Duffy 1999). As possible explanation for this difference we recall that our analytical solution is based on the two continuity conditions (\ref{f300}) and (\ref{f4}) at the shock. These two continuity  conditions are needed as our steady-state diffusion-convection transport equation (\ref{f1}) is a second-order differential equation in the position coordinate $z$. While the continuity condition (\ref{f300}) for the particle phase density at the shock is also used in the eigenfunction solution method, the continuity condition (\ref{f4}) for the flux of particles is not used in that method as the \fo transport equation (\ref{a3}) is a first-order differential equation in the position coordinate $z$. It is clear that the use of different continuity 
conditions results in different results. 
\section {Summary and conclusions}
The analytical theory of diffusive cosmic ray acceleration at parallel stationary shock waves with magnetostatic turbulence is generalized to arbitrary shock speeds $V_s=\beta _1c$, including in particular relativistic speeds. This is achieved by applying the diffusion approximation to the relevant Fokker-Planck particle transport equation formulated in the mixed comoving coordinate system. In this coordinate system the particle's momentum coordinates $p$ and $\mu =p_{\parallel }/p$ are taken in the rest frame of the streaming plasma, whereas the time and space coordinates are taken in the observer's system. 
The \fo particle transport equation contains connection coefficients resulting from the coordinate transformations into this mixed frame which are properly included in the diffusion approximation. For magnetostatic slab turbulence the diffusion-convection transport equation (\ref{d20}) for the isotropic (in the rest frame of the streaming plasma) part of the particle's phase space density is derived for the first time for arbitrary shock speeds. 
In the limit of nonrelativistic flows the diffusion-convection transport equation differs from the transport equation used in earlier nonrelativistic diffusive shock acceleration theory (Axford et al. 1977, Krymsky 1987, Blandford and Ostriker 1978, Bell 1978, Drury 1983) by an additional term. This results from our correct handling of the connection coefficients. The additional term implies a velocity dependence of the acceleration rate and thus provides a major modification of the resulting differential momentum spectrum of accelerated particles in the nonrelativistic flow limit at nonrelativistic particles momenta: instead of a power law distribution of accelerated particles at the shock a Lorentzian distribution function results, which at large momenta then approaches the power law distribution inferred in earlier acceleration theories for nonrelativistic shock speeds. 

For a step-wise shock velocity profile the steady-state diffusion-convection transport equation is solved analytically again for the first time for arbitrary shock speeds following closely the solution method developed for nonrelativistic speeds, making use of the continuity conditions (\ref{f300}) and (\ref{f4}) for the \kr phase space density and streaming density at the shock. For a symmetric pitch-angle scattering Fokker-Planck coefficient $D_{\mu \mu }(-\mu )=D_{\mu \mu }(\mu )$  the steady-state solution is independent of the microphysical scattering details. For nonrelativistic mono-momentum particle injection at the shock the differential number density of accelerated particles is a Lorentzian-type distribution function which at large momenta approaches a power law distribution function $N(p\ge p_c)\propto p^{-\xi }$ with the spectral index $\xi (\beta _1) =1+[3/(\Gamma _1\sqrt{r^2-\beta _1^2}-1)(1+3\beta _1^2)]$. For nonrelativistic ($\beta _1\ll 1$) shock speeds this spectral index agrees with the 
known result $\xi (\beta _1\ll 1)\simeq (r+2)/(r-1)$, whereas for ultrarelativistic ($\Gamma _1\gg 1$) shock speeds the spectral index value is close to unity. If particle injection occurs already at relativistic momenta, 
the steady-state solution is of power law type at all higher particle momenta. 

For asymmetric pitch-angle scattering Fokker-Planck coefficient $D_{\mu \mu }(-\mu )\ne D_{\mu \mu }(\mu )$, resulting from magnetostatic isospectral slab Alfven waves with non-zero values of the magnetic and cross helicities, the momentum spectrum of accelerated particles depends on the microphysical details of particle's pitch angle scattering. In particular, a dependence of the momentum spectrum on the charge sign of the \kr particles is found. 
For turbulence spectral indices $s<2$, however, this difference is negligibly small. But steeper turbulence power spectra with $s\ge 2$ may provide stronger 
differences, which needs to be investigated in the future. Further future research topics will be concerned with the derivation of time-dependent solutions of the generalized diffusion-convection transport equation, the study of non-stationary flows, the inclusion of finite up- and downstream free escape boundary conditions, the investigation of the influence of finite frequency effects for low-frequency Alfven waves discarding the magnetostatic approximation, and the investigation of the influence of \kr momentum losses. These additional effects have been studied before in great detail for nonrelativistic shocks, and it is of high interest to investigate their importance for relativistic shocks. The work presented here provides the analytical basis for these future studies. 
\acknowledgements
I thank Msc Steffen Krakau and MSc Thorsten Antecki for help with the figures and a critical reading of the manuscript. This work was partially supported by the Deutsche Forschungsgemeinschaft through grants Schl 201/23-1 and Schl 201/29-1.

\section {Appendix A: the integral $I(p,p_0)$}
The integral (\ref{f26}) is be solved in closed form by

\bdm
I(\beta ,\beta _0)=w(\beta )-w(\beta _0),\;\;\; w(x)={W_1(x)\over W_2},
\eba
W_1(x)=2(1-3\beta _1^2)\arctan \left({ZR\beta _1+2x\over \beta _1\sqrt{12-R^2}}\right)
\eba
+\sqrt{12-R^2}\Bigl[\left(ZR\beta _1-1-3\beta _1^2\right)\ln (x-1)
\eba
-\left(1+3\beta _1^2+ZR\beta _1\right)\ln (1+x)
\eba
+\left(1+3\beta _1^2\right)\ln (3\beta _1^2+ZR\beta _1x+x^2)\Bigr],
\ebe
W_2=2\left[\left(1+3\beta _1^2\right)^2-R^2\beta _1^2\right)\sqrt{12-R^2}
\label{j1}
\ee

\end{document}